\documentclass[lettersize,journal]{IEEEtran}
\usepackage{amsmath,amsfonts}
\usepackage{yhmath}
\usepackage{algorithmic}
\usepackage{algorithm}
\usepackage{array}
\usepackage[caption=false,font=normalsize,labelfont=sf,textfont=sf]{subfig}
\usepackage{textcomp}
\usepackage{stfloats}
\usepackage{url}
\usepackage{verbatim}
\usepackage{graphicx}
\usepackage{cite}
\begin{document}

\title{Spin and Orbital Angular Momenta of Electromagnetic Waves: From Classical to Quantum Forms}

\author{Wei E. I. Sha, \textit{Senior Member, IEEE}, Zhihao Lan, Menglin L. N. Chen, \textit{Senior Member, IEEE}, Yongpin P. Chen, \textit{Senior Member, IEEE}, Sheng Sun, \textit{Senior Member, IEEE}
\thanks{
This work was supported by the National Natural Science Foundation of China under Grant 61975177 and Grant U20A20164.  (Corresponding author: Wei E. I. Sha)
	
W. E. I. Sha is with the Key Laboratory of Micro-Nano Electronic Devices and Smart Systems of Zhejiang Province, College of Information Science and Electronic Engineering, Zhejiang University, Hangzhou 310027, China (email: weisha@zju.edu.cn).

Z. Lan is with the Department of Electronic and Electrical Engineering, University College London, London WC1E 6BT, United Kingdom.

M. L. N. Chen is with Department of Electrical and Electronic Engineering, The Hong Kong Polytechnic University, Hong Kong, China.

Y. P. Chen and S. Sun are with School of Electronic Science and Engineering, University of Electronic Science and Technology of China, Chengdu 611731, China.
}

}

\markboth{}%
{Shell \MakeLowercase{\textit{et al.}}: A Sample Article Using IEEEtran.cls for IEEE Journals}


\maketitle

\begin{abstract}
Angular momenta of electromagnetic waves are important both in concepts and applications. In this work, we systematically discuss two types of angular momenta, i.e., spin angular momentum and orbital angular momentum in various cases, e.g., with source and without source, in classical and quantum forms. Numerical results demonstrating how to extract the topological charge of a classical vortex beam by spectral method are also presented.
\end{abstract}


\section{Introduction}
\IEEEPARstart{E}{lectromagnetic} (EM) waves carry two types of angular momentum, namely, spin angular momentum (SAM) and orbital angular momentum (OAM)\cite{Bliokh13NJP}. Different from SAM which provides only two eigenstates; by engaging OAM, we can construct unlimited number of eigenstates and each of them is identified by the OAM index $l$. In 1992, Allen et al first found that Laguerre-Gaussian (LG) beams carry well defined OAM \cite{Allen92PRA}. Since then, OAM beams have been widely explored in optics. Based on the light-matter interaction, the unique features of OAM beams make them irreplaceable in many applications, such as optical tweezers \cite{Simpson97OL}, sensors \cite{Lavery13Science} and so on. More importantly, by utilizing the orthogonality of different OAM modes, EM waves carrying OAM provide an alternative and innovative platform for spatial multiplexing of wireless communication systems to potentially increase channel capacity \cite{Wang12NPho,Willner15AOP,Yan14NC,Yuan21PRApp}. From optical to radio regimes, different techniques have been adopted to generate and detect OAM, such as waveplates \cite{Marrucci06PRL}, metasurfaces \cite{Chen16JAP,Chen17TAP} and antenna arrays \cite{Thide07PRL}.

Although several pieces of theoretical works on SAM and OAM of EM waves have been reported \cite{Bliokh13NJP,AM_Yao,KY_Bliokh,Barnett16JO,Bliokh15PhysRep,Yang22PRResearch}, there are still challenges to learn and understand these theories by researchers in engineering EM society. First, the current theories and calculations about these topics are not well organized. Most studies only look at one part of the theory, leaving out other important aspects. This makes it hard to get a full picture of how SAM and OAM work in EM waves.  Second, the formulae used to describe these phenomena are complex. This complexity makes it difficult for researchers, especially those new to this field, to understand these concepts clearly.  Third, there's a lack of connection between classical physics theories and quantum physics theories in this area. Discussions comparing SAM and OAM of photons (light particles) and electrons (particles in atoms) are also rare. Finally, because of above issues, we need to review and re-examine the theories of SAM and OAM of EM waves. This is not just important for research, but also for teaching these concepts in EM science and engineering.

\section{Theory}
\subsection{Field momentum}

\begin{figure}[!t]
\centering
\includegraphics[width=\columnwidth]{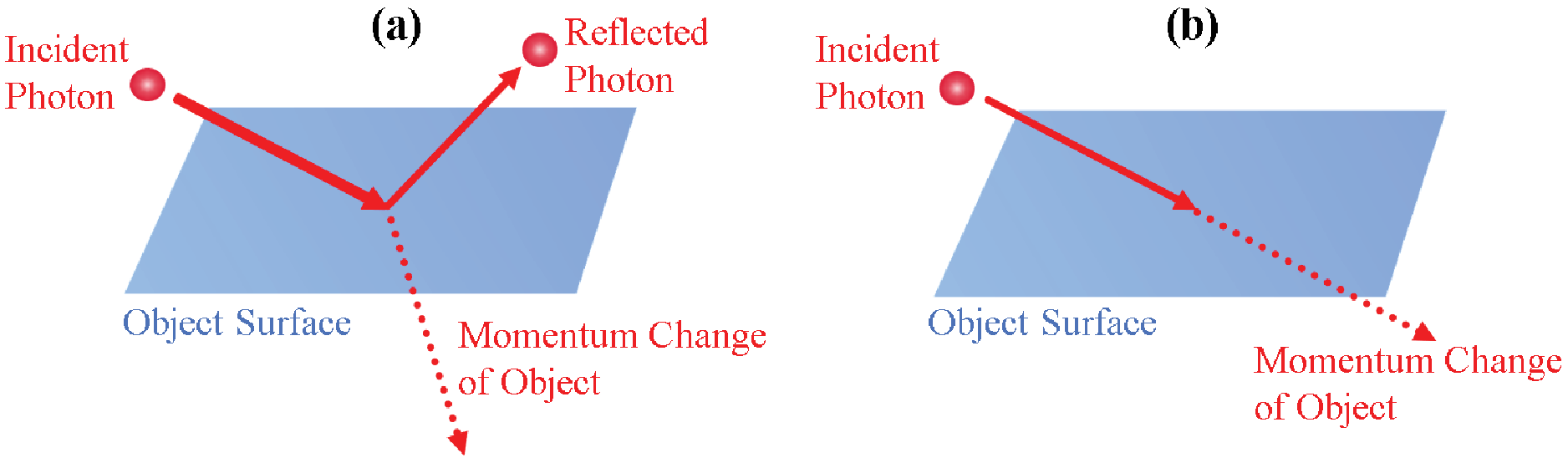}
\caption{A photon incident on an opaque object surface will be (a) reflected back or (b) absorbed, transferring its momentum to the object. }
\label{fig_1}
\end{figure}

A photon will be reflected back or absorbed, when it is incident on an opaque object. Consequently, the photon will transfer its momentum to the object (see Fig. \ref{fig_1}). From De Broglie hypothesis, photon momentum is given by
\begin{equation}
p=\hbar k = \hbar \frac{\omega}{c}=\frac{E}{c}
\label{Eq1}
\end{equation}
where $\hbar$ is the reduced Planck constant, $k$ is the wavenumber, $\omega$ is the angular frequency, and $c$ is the speed of light. From wave-particle duality, photon energy $E$ can be converted to classical field form in free space, i.e.
\begin{gather}
\mathbf{P}^f = \int_v \frac{\mathbf{E} (\mathbf{r},t)\times \mathbf{H} (\mathbf{r},t) dsdt}{c}
= \int_v \frac{\mathbf{E} (\mathbf{r},t)\times \mathbf{H} (\mathbf{r},t)}{c}\frac{dsdl}{c} \nonumber \\
=\int_v \frac{\mathbf{E} (\mathbf{r},t)\times \mathbf{H} (\mathbf{r},t)}{c^2} dv =  \int_v \mathbf{p}^f(\mathbf{r},t) dv
\label{Eq2}
\end{gather}
where $ds$ ($dl$) is the differential area (length) that is transverse (parallel) to the propagation direction of EM waves; and $dt=dl/c$ is the differential time. Moreover, $\mathbf{E}$ and $\mathbf{H}$ are the electric and magnetic fields; and $ \mathbf{p}^f$ and $ \mathbf{P}^f$ are the field momentum density and field momentum, where the superscript $f$ denotes the ``field".

The summation of the field and mechanical momenta is canonical momentum. The time varying of the canonical momentum is related to EM or optical force expressed as Maxwell's stress tensor \cite{Jackson99book}.

\begin{gather}
\frac{\partial \mathbf{P}}{\partial t} = \frac{\partial \mathbf{P}^f}{\partial t} +\frac{\partial \mathbf{P}^m}{\partial t} = \oint_s \mathbf{\bar{T}}\cdot d\mathbf{s}=\mathbf{F} \\
\frac{\partial \mathbf{P}^m}{\partial t} = q \mathbf{E}+q\mathbf{v}\times \mathbf{B} \\
\mathbf{\bar{T}}=\epsilon_0\mathbf{EE}-\frac{1}{2}\epsilon_0 |\mathbf{E}|^2\mathbf{\bar{I}}+\mu_0\mathbf{HH}-\frac{1}{2}\mu_0 |\mathbf{H}|^2\mathbf{\bar{I}}
\label{Eq3}
\end{gather}
where $\mathbf{P}$ is the canonical momentum; $\mathbf{P}^m$ is the mechanical momentum and $\partial  \mathbf{P}^m /\partial t$ gets the Lorentz force, where the superscript $m$ denotes the ``mechanical". Moreover, $\mathbf{\bar{T}}$ is the Maxwell's stress tensor and the optical force $\mathbf{F}$ can be obtained by a closed surface integral $\oint_s \mathbf{\bar{T}}\cdot d\mathbf{s}$ surrounding an object. Furthermore, $q$ is the elementary charge, $\mathbf{B}$ is the magnetic flux density, and $\epsilon_0$ and $\mu_0$ are the permittivity and permeability of free space, respectively. Additionally, $\mathbf{EE}$ and $\mathbf{HH}$ denote tensor products of EM fields, and $\mathbf{\bar{I}}$ is a $3\times3$ identity tensor.

\subsection{Field momentum at source region}
At source region with moving charges, the near-field electric field is longitudinal ($ \mathbf{E}^{\parallel}$) and the field momentum can be simplified as
\begin{gather}
 \mathbf{P}^f=\epsilon_0 \int_v \mathbf{E}^{\parallel} \times \mathbf{B} dv=\epsilon_0 \int_v - \nabla\phi \times \nabla \times \mathbf{A}dv \nonumber \\
  =\int_v - \epsilon_0 \nabla^2\phi  \mathbf{A}dv = \int_v\rho   \mathbf{A}dv
\end{gather}
where $\phi$ and $\mathbf{A}$ are the scalar and vector potentials, respectively. The integral identity is easily obtained with the help of Einstein summation notation and integration by part. For one moving charged particle in EM fields, (6) is nothing but field momentum $q\mathbf{A}$.

When considering charge-field interaction, the mechanical momentum of the free charged particle can be obtained by substracting the field momentum $q\mathbf{A}$ from the canonical momentum $\mathbf{P}$. Consequently, the corresponding classical Hamiltonian can be constructed as,
\begin{gather}
H=\frac{(\mathbf{P}-q \mathbf{A} (\mathbf{r},t))^2}{2m}+q\phi(\mathbf{r},t)
\end{gather}
where $m$ is the mass of the particle; and the first and second terms are the kinetic and potential energy of the free charged particle, respectively. Furthermore, the canonical momentum will be quantized for the quantum form of the classical Hamiltonian above, which is used to simulate Rabi oscillation in semiclassical atom-field interaction \cite{Chen17CPC,Xie19JMMCT,Ryu16JMMCT}.
\begin{gather}
\hat{H}=\frac{(\mathbf{\hat{P}}-q \mathbf{A} (\mathbf{r},t))^2}{2m}+q\phi(\mathbf{r},t)+\hat{V}(\mathbf{r})
\end{gather}
where $\mathbf{\hat{P}}=-i\hbar \nabla$ is the momentum operator in quantum mechanics and $\hat{V}$ is the quantum-confinement potential operator of the (artificial) atom. It is worth noting that a physical quantity with the overhat notation means an operator and does not correspond to a deterministic value.

\subsection{Field angular momentum at sourceless region}
At sourceless region without charges, the electric field is transverse ($\mathbf{E}^{\perp}$) and the field momentum is given by
\begin{gather}
\mathbf{P}^f =  \int_v \mathbf{p}^fdv = \int_v\epsilon_0  \mathbf{E}^{\perp}\times \mathbf{B} dv
\end{gather}
Using $\mathbf{B}=\nabla \times \mathbf{A}$ and the expression of field angular momentum density $\mathbf{j}^f=\mathbf{r} \times \mathbf{p}^f$, the field angular momentum $\mathbf{J}^f$ can be written as \cite{Garrison08book}
\begin{gather}
\mathbf{J}^f=\int_v \mathbf{j}^f dv=\int_v\epsilon_0  \mathbf{r} \times \mathbf{E}^{\perp}\times \nabla \times \mathbf{A} dv \nonumber \\
=\epsilon_0\int_v\sum_j E_j^{\perp}(\mathbf{r}\times \nabla)A_j dv +\epsilon_0 \int_v\mathbf{E}^{\perp}\times \mathbf{A} dv \nonumber  \\
=\mathbf{L}^f+\mathbf{S}^f
\end{gather}
where the first term ($\mathbf{L}^f$) and the second term ($\mathbf{S}^f$) are the OAM and SAM of EM fields, respectively. The above can be derived by using the vector identity $\mathbf{F}\times(\nabla\times\mathbf{G})=\sum_j F_j \nabla G_j-(\mathbf{F}\cdot\nabla)\mathbf{G}$ and integration by part with the divergence-free condition $\nabla\cdot\mathbf{E}^{\perp}=0$. The OAM depends on the field gradient, where inhomogeneous field structure plays a role. The SAM depends on the polarization of fields, and will get zero (non-zero) for linear (circular) polarized plane waves.

For time-harmonic fields in free space with the time convention of $\text{exp}(-i\omega t)$, we can split $\mathbf{E}$ and $\mathbf{A}$ fields into their positive- and negative-frequency components
\begin{gather}
\mathbf{E}^{\perp}(\omega,\mathbf{r})=\mathbf{E}^{\perp,+}(\omega,\mathbf{r})+\mathbf{E}^{\perp,-}(\omega,\mathbf{r}), \nonumber \\
\mathbf{A}^{\perp}(\omega,\mathbf{r})=-i\frac{1}{\omega} \left [ \mathbf{E}^{\perp,+}(\omega,\mathbf{r})-\mathbf{E}^{\perp,-}(\omega,\mathbf{r}) \right]
\end{gather}
where $\mathbf{E}^{\perp}=-\partial\mathbf{A}^{\perp}/\partial t$ with the radiation gauge and $\mathbf{E}^{\perp,+}=\left (\mathbf{E}^{\perp,-}\right)^*$. Taking a plane wave as an example, $\mathbf{E}^{\perp,+}(\omega,\mathbf{r})=\exp(-i\omega t+i\mathbf{k}\cdot \mathbf{r})$ and $\mathbf{k}$ is the wavevector.

By using (9-11), the time-averaged field momentum, OAM and SAM are of the following forms
\begin{gather}
\mathbf{P}^f =\hbar\omega\mu_0\int_v \left( \sqrt{\frac{2\epsilon_0}{\hbar\omega}}\mathbf{{E}}^{\perp,+} \right)^*\times \left(\sqrt{\frac{2\epsilon_0}{\hbar\omega}}\mathbf{{H}}^{\perp,+}\right) dv \nonumber \\
=\hbar\omega\mu_0\int_v \left( \mathbf{\tilde{E}}^{\perp,+} \right)^*\times \left(\mathbf{\tilde{H}}^{\perp,+}\right) dv\\
\mathbf{L}^f=\int_v\sum_j \left ( \sqrt{\frac{2\epsilon_0}{\hbar\omega}}E_j^{\perp,+}\right)^* (\mathbf{r}\times -i\hbar \nabla)  \left ( \sqrt{\frac{2\epsilon_0}{\hbar\omega}}E_j^{\perp,+}\right) dv \nonumber \\
=\int_v\sum_j \left( \tilde{E}_j^{\perp,+} \right)^* \mathbf{\hat{L}}\left( \tilde{E}_j^{\perp,+} \right) dv \\
\mathbf{S}^f=\int_v(-i\hbar) \left ( \sqrt{\frac{2\epsilon_0}{\hbar\omega}}\mathbf{E}^{\perp,+}\right)^*\times  \left ( \sqrt{\frac{2\epsilon_0}{\hbar\omega}}\mathbf{E}^{\perp,+}\right) dv \nonumber \\
=\int_v \left( \mathbf{\tilde{E}}^{\perp,+} \right)^* \mathbf{\hat{S}}^f\left( \mathbf{\tilde{E}}^{\perp,+} \right) dv
\end{gather}
where the scaled factor $(2\epsilon_0/\hbar\omega)^{1/2}$ is connected to the canonical quantization of Maxwell's equations in free space \cite{Scully97book}. Moreover, $\mathbf{\hat{L}}=\mathbf{r}\times \mathbf{\hat{P}}$ is the angular momentum operator that is a Hermitian operator, and $\mathbf{\hat{S}}^f$ are the $3\times3$ Pauli matrices of photon, which are given by
\begin{gather}
\mathbf{\hat{S}}_x^f=-i\hbar
\begin{pmatrix}
0 & 0 &0 \\
0 & 0 &1 \\
0 & -1 &0
\end{pmatrix},\,
\mathbf{\hat{S}}_y^f=-i\hbar
\begin{pmatrix}
0 & 0 &-1 \\
0 & 0 &0 \\
1 & 0 &0
\end{pmatrix}, \nonumber \,\\
\mathbf{\hat{S}}_z^f=-i\hbar
\begin{pmatrix}
0 & 1 &0 \\
-1 & 0 &0 \\
0 & 0 &0
\end{pmatrix}
\end{gather}

The mathematical expressions of OAM and SAM of EM fields (13-14) show a beautiful analogy to those of electron (16), i.e.

\begin{gather}
L^e=\int_v \Psi^*\mathbf{\hat{L}}\Psi dv, \hspace{0.2cm}S^e=\int_v \Psi^*\mathbf{\hat{S}}^e\Psi dv
\end{gather}
where $\Psi$ is the wave function of electron, $\mathbf{\hat{L}}$ is the same angular momentum operator, and $\mathbf{\hat{S}}^e$ are the $2\times2$ Pauli matrices of electron with the same commutation relations $[\mathbf{\hat{S}}_x^{e},\mathbf{\hat{S}}_y^{e}]=i\hbar\mathbf{\hat{S}}_z^{e}$ and $[\mathbf{\hat{S}}_x^{f},\mathbf{\hat{S}}_y^{f}]=i\hbar\mathbf{\hat{S}}_z^{f}$. The different dimensions of the photon and electron Pauli matrices are due to the fact that photon is the boson with spin of $1$ and electron is the fermion with spin of $1/2$.

For a classical vortex beam carrying SAM and OAM and propagating along the $z$ direction with the wavenumber of $k_z$, under the paraxial approximation, Laguerre-Gaussian modes for the scaled $\mathbf{E}$ field in cylindrical coordinates $(\rho, \phi, z)$ are
\begin{gather}
\mathbf{\tilde{E}}^{\perp, +}(\rho, \phi, z; l)=\mathbf{e}_{\sigma}F^{|l|}(\rho,z)\text{exp}(-i\omega t +il\phi+ik_z z)
\end{gather}
where $F^{|l|}(\rho,z)$ are the functions for representing the Laguerre-Gaussian modes, $l=\pm1,\pm2,\cdots$ and $\sigma=\pm1$ are called the orbital and spin topological charges, and $\mathbf{e}_{\sigma}$ are the left $\left[(\mathbf{e}_x+i\mathbf{e}_y)/\sqrt{2}\right]$ and right $\left[(\mathbf{e}_x-i\mathbf{e}_y)/\sqrt{2}\right]$ circularly polarized unit vectors. Then, in view of (13-14), $\mathbf{r}\rightarrow\mathbf{\rho}$ for the paraxial beam in cylindrical coordinates and orthonormal property of the eigenmodes $\langle\mathbf{\tilde{E}}^{\perp, +}(\mathbf{r};l),\mathbf{\tilde{E}}^{\perp, +}(\mathbf{r};l')\rangle=\delta_{ll'}$, we can easily get
\begin{gather}
L_z^f=\hbar l, \hspace{0.2cm} \mathbf{S}^f=\hbar\sigma\mathbf{e}_z
\end{gather}
where we only consider a single eigenmode.

\subsection{Quantization of field angular momentum }
Using the canonical quantization method of free EM fields, the positive-frequency components of the scaled $\mathbf{E}$ and $\mathbf{H}$ fields can be quantized as \cite{Scully97book}
\begin{gather}
\mathbf{\hat{\tilde{E}}}^{\perp,+}=\sum_{\mathbf{k},\sigma}\mathbf{e}_{\sigma}\hat{a}_{\mathbf{k},\sigma} \text{exp}(-i\omega_kt+i\mathbf{k}\cdot\mathbf{r}) \\
\mathbf{\hat{\tilde{H}}}^{\perp,+}=\sum_{\mathbf{k},\sigma}\frac{\mathbf{k}\times\mathbf{e}_{\sigma}}{\omega_k\mu_0}\hat{a}_{\mathbf{k},\sigma} \text{exp}(-i\omega_kt+i\mathbf{k}\cdot\mathbf{r})
\end{gather}
where $\hat{a}_{\mathbf{k},\sigma}$ is the annihilation operator of photon, and the summation is for all eigenmodes with left or right circular polarization.

Using (12) and (19-20), the quantized field momentum arrives at such an elegant expression
\begin{gather}
\mathbf{\hat{P}}^f=\sum_{\mathbf{k},\sigma}\hbar \mathbf{k} \hat{a}_{\mathbf{k},\sigma}^{\dagger}\hat{a}_{\mathbf{k},\sigma}
\end{gather}
where $\hat{a}_{\mathbf{k},\sigma}^{\dagger}$ is the creation operator of photon, non-commutable with the annihilation operator $\hat{a}_{\mathbf{k},\sigma}$. The two operators have the commutation relation $[\hat{a}_{\mathbf{k},\sigma},\hat{a}_{\mathbf{k},\sigma}^{\dagger}]=1$. Moreover, the normal ordering form $\hat{a}_{\mathbf{k},\sigma}^{\dagger}\hat{a}_{\mathbf{k},\sigma}$ is used in (21) to describe photon absorption, which is different from the anti-normal ordering form $\hat{a}_{\mathbf{k},\sigma}\hat{a}_{\mathbf{k},\sigma}^{\dagger}$ describing photon emission.

Using (14) and (19), the quantized SAM can be obtained as
\begin{gather}
\mathbf{\hat{S}}^f=\sum_{\mathbf{k},\sigma}\hbar \mathbf{e}_{\mathbf{k}} \sigma \hat{a}_{\mathbf{k},\sigma}^{\dagger}\hat{a}_{\mathbf{k},\sigma}
\end{gather}
where $\mathbf{e}_{\mathbf{k}}$ is the unit vector along the $\mathbf{k}$ direction.

By using properties of Fourier transform $\mathbf{r}\leftrightarrow i\nabla_{\mathbf{k}}$ and $\nabla \leftrightarrow i \mathbf{k}$, the classical OAM expression (13) can be quantized as an integral in the $\mathbf{k}$ space.
\begin{gather}
\mathbf{\hat{L}}^f=\frac{-i\hbar}{(2\pi)^3}\int_v\sum_j\left(\hat{M}_j(\mathbf{k})\right)^*\left( \mathbf{k} \times \frac{\partial}{\partial \mathbf{k}} \right) {\hat{M}}_j(\mathbf{k})d\mathbf{k}
\end{gather}
where $\mathbf{\hat{M}}(\mathbf{k})=\sum_{\mathbf{k},\sigma}\mathbf{e}_{\sigma}\hat{a}_{\mathbf{k},\sigma}$, which is the same as the mathematical description in the book written by Garrison and Chiao (see equation (3.55) in \cite{Garrison08book}). Particularly, if the eigenmodes (17) are employed, the $z$ component of the quantized OAM is
\begin{gather}
\hat{L}^f_z=\sum_{\mathbf{k},\sigma}\hbar l \hat{a}_{\mathbf{k},\sigma}^{\dagger}\hat{a}_{\mathbf{k},\sigma}
\end{gather}

Regarding (22) and (24), $\hat{a}_{\mathbf{k},\sigma}^{\dagger}\hat{a}_{\mathbf{k},\sigma}$ represents the photon number operator; Thus, the quantized SAM and OAM exactly are the analogy to their classical counterparts (18), where each eigenmode is quantized as a quantum harmonic oscillator \cite{WC_Chew}.

Quantized OAM states have the potential to construct high-dimensional quantum systems with information-processing capability better than two-level single-photon systems. Recently, a pair of entangled states of OAM and a superposition state of OAM have been experimentally generated from single-photon sources \cite{Cuo_Wu,Bo_Chen}. Thus, more information can be encoded on each photon allowing for several emerging applications, such as high-dimensional quantum key distribution and quantum computing, super-resolution quantum imaging, and high-precision quantum sensors.

\section{Numerical Results}
In this section, we will show how to numerically extract the topological charge of a classical vortex beam. By using (13) and (18), the topological charge of the paraxial vortex beam is given by
\begin{gather}
l=\int_s\sum_j \left(  \tilde{E}_j^{\perp,+}\right)^* -i\frac{\partial}{\partial \phi}\left(  \tilde{E}_j^{\perp,+}\right) d{s}
\end{gather}
This formula is related to the Berry connection and Berry phase of vortex beams \cite{KY_Bliokh2,Gangaraj17JMMCT}. For the paraxial vortex beam propagating along the $z$ direction, the summation index $j$ is only for the $x$ and $y$ polarized components.

Moreover, the scaled $\tilde{E}_j^{\perp,+}$ should be normalized as
\begin{gather}
\int_s \left(\tilde{E}_j^{\perp,+}\right)^* \left(  \tilde{E}_j^{\perp,+}\right) d{s}=1
\end{gather}
The derivative operation  $\frac{\partial}{\partial \phi}=x\frac{\partial}{\partial y}-y\frac{\partial}{\partial x} $ can be discretized with the spectral method and fast Fourier transform \cite{Liu99TGRS}.

\begin{figure}[!t]
\centering
\includegraphics[width=\columnwidth]{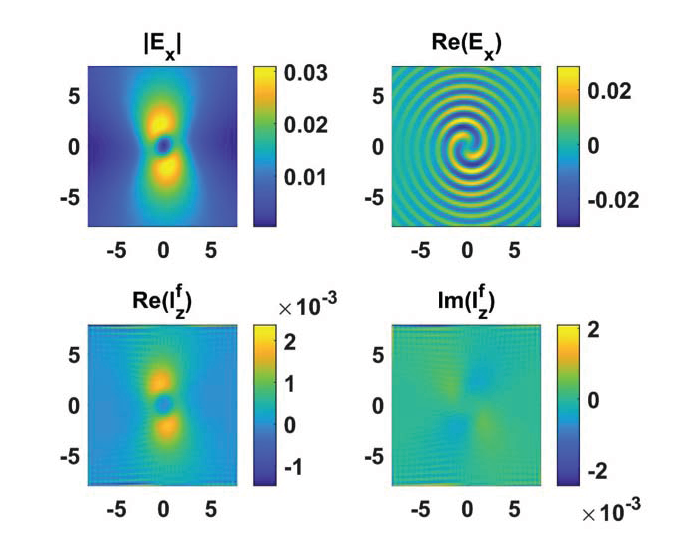}
\caption{Spatial distributions of electric field and OAM density for a vortex beam with the topological charge of $l=2$, (a) amplitude of $E_x$ field; (b) real part of $E_x$ field; (c) real part of $z$ component of OAM density; (d) imaginary part of $z$ component of OAM density.}
\label{fig_2}
\end{figure}

\begin{figure}[!t]
\centering
\includegraphics[width=\columnwidth]{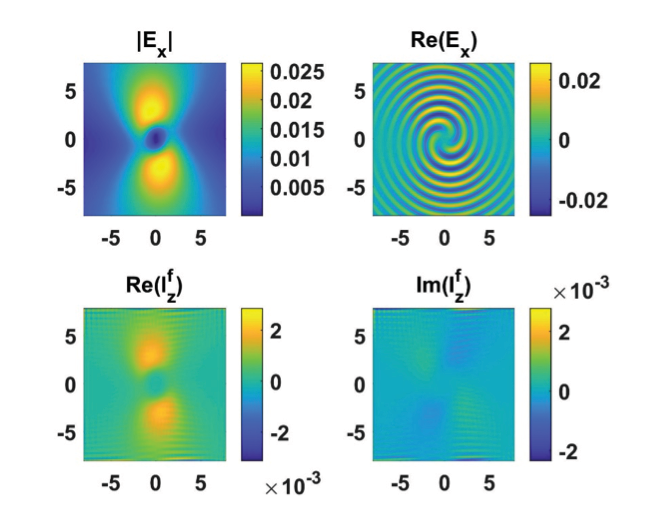}
\caption{Spatial distributions of electric field and OAM density for a vortex beam with the topological charge of $l=3$, (a) amplitude of $E_x$ field; (b) real part of $E_x$ field; (c) real part of $z$ component of OAM density; (d) imaginary part of $z$ component of OAM density.}
\label{fig_3}
\end{figure}

We consider $12$ $x$-polarized Hertzian dipoles uniformly distributed along a circle of radius of $0.5\lambda$ at a source plane perpendicular to the $z$ direction, where $\lambda$ is the incident wavelength. Phase distributions of the dipole sources are set to be  $\text{exp}(il\phi)$ for generating a classical vortex beam propagating along the $z$ direction with the topological charge of $l$. The observation plane with the size of $16\lambda$ $\times$ $16\lambda$ is $2\lambda$ above the source plane. Figures \ref{fig_2} and \ref{fig_3} show the $x$ component of electric fields and $z$ component of OAM densities for $l=2$ and $l=3$ vortex beams, respectively. The spatial resolution is $0.2\lambda$. Moreover, the calculated topological charges are listed in Table \ref{tab_1}. Here, zero padding is employed to avoid aliasing and minimize edge effects. The spectral method could get reasonably good results even with a coarse resolution of $0.4\lambda$.

In fact, a classical vortex beam has an integer topological charge as the topological invariant of propagation. The spectral method is a way to calculate the topological invariant, which is as important as the Chern number calculation in topological insulators \cite{Ran_Zhao}.

\begin{table}
\centering
\caption{Calculated topological charges $l$ of vortex beams with different spatial resolutions $\Delta$.  }
\begin{tabular}{|p{1.5cm}|p{1.5cm}|p{1.5cm}|p{1.5cm}|}
    \hline
    $l$/$\Delta$ & $0.4$$\lambda$ & $0.3$$\lambda$ &$0.2$$\lambda$ \\
    \hline
    $l=3$ & $2.959$ & $2.977$ & $2.990$ \\
    \hline
    $l=2$ & $1.987$ & $1.986$ & $1.996$ \\
    \hline
\end{tabular}
\label{tab_1}
\end{table}

\section{Conclusion}

In summary, we have systematically discussed the SAM and OAM of EM waves. In particular, the canonical momentum is the summation of the field and mechanical momenta; and its time varying is related to EM or optical force expressed as the Maxwell's stress tensor. At source regions with moving charges, the canonical momentum could be used to construct a classical Hamiltonian describing the charge-field interaction. Moreover, the canonical momentum will be quantized for the quantum form of the classical Hamiltonian to model the semiclassical atom-field interaction. At sourceless region without charges, the field angular momentum could be decomposed as the SAM and OAM, which show the analogy to those of electron. Under the paraxial approximation, a classical vortex beam propagating along the $z$ direction can carry both SAM and OAM in the $z$ direction. While the SAM depends on the polarization of fields and thus can only take two values $(\sigma=\pm1)$; the OAM depends on the field gradient and its topological charge can take any integers ($l=\pm1,\pm2,\cdots$). For quantization of SAM and OAM, both of the quantized forms can be elegantly expressed as the photon number operator with the topological charge as a proportional constant. Finally, numerical results showing how to extract the topological charge of a classical vortex beam are presented. This work can serve as a useful guide on studying SAM and OAM of EM waves.

In future, several unsolved or emerging problems in this area could be explored. On one hand, mathematical descriptions of SAM and OAM have been controversial due to the issues of gauge invariance and Lorentz covariance, e.g., the total angular momentum is separated into spin and orbital parts, however the results are usually not gauge invariant \cite{Barnett16JO, H_Chen}.  On the other hand, while the separation of angular momentum into longitudinal (i.e., aligned with the propagation direction) spin and orbital parts is more or less straightforward in paraxial monochromatic beams, the situation is more subtle in generic non-paraxial or non-monochromatic fields, where transverse SAM and OAM could exist in various structured fields including evanescent waves, interference fields, and focused beams \cite{Bliokh15PhysRep}. There remain many intriguing theories that merit further investigation.

\end{document}